\documentclass{aastex}
\usepackage{spr-astr-addons}
\usepackage{url}

\RequirePackage{color}

\newcommand{\be}{\begin{equation}}
\newcommand{\ee}{\end{equation}}
\newcommand{\bea}{\begin{eqnarray}}
\newcommand{\eea}{\end{eqnarray}}
\newcommand{\beaa}{\begin{eqnarray*}}
\newcommand{\eeaa}{\end{eqnarray*}}

\newcommand{\nn}{\nonumber \\}
\newcommand{\e}{\mathrm{e}}

\def\be{\begin{equation}}
\def\ee{\end{equation}}
\def\bea{\begin{eqnarray}}
\def\eea{\end{eqnarray}}

\def\nn{\nonumber \\}
\def\e{\mathrm{e}}

\begin{document}

\title{Phantom without ghost}
\slugcomment{Not to appear in Nonlearned J., 45.}
\shorttitle{Short article title}
\shortauthors{Autors et al.}

\author{Shin'ichi Nojiri\altaffilmark{1,2}} \and \author{Emmanuel
N.~Saridakis\altaffilmark{3,4}}

\altaffiltext{1}{Department of Physics, Nagoya University, Nagoya 
464-8602,
Japan.}
\altaffiltext{2}{ Kobayashi-Maskawa Institute for the Origin of Particles and
the Universe, Nagoya University, Nagoya 464-8602, Japan.}
\altaffiltext{3}{Physics Division, National Technical University of Athens,
15780 Zografou Campus, Athens, Greece.}
\altaffiltext{4}{Instituto de
F\'{\i}sica, Pontificia Universidad Cat\'olica de Valpara\'{\i}so,
Casilla 4950, Valpara\'{\i}so, Chile.}

\begin{abstract}
The Nine-Year WMAP results combined with other cosmological data seem to
indicate an enhanced favor for the phantom regime, comparing to previous
analyses. This behavior, unless reversed by future observational data,
suggests to consider the phantom regime more thoroughly. In this work we
provide three modified gravitational scenarios in which we obtain the
phantom realization without the appearance of ghosts degrees of freedom,
which plague the naive approaches on the subject, namely the Brans-Dicke
type gravity, the scalar-Einstein-Gauss-Bonnet gravity, and the $F(R)$
gravity, which are moreover free of perturbative instabilities.
The phantom regime seems to favor the gravitational modification instead of
the universe-content alteration.
\end{abstract}

\keywords{Phantom cosmology; Modified gravity; Dark energy; Ghost
instabilities.}


\section{Introduction}

The recently announced Nine-Year Wilkinson Microwave Anisotropy Probe
(WMAP) results \citep{Hinshaw:2012fq} indicate that
the Equation of State
(EoS) parameter of the dark energy $w_\mathrm{DE}$, which is defined by the
ratio of the pressure $p_\mathrm{DE}$ and the energy density
$\rho_\mathrm{DE}$ of the dark energy, $w_\mathrm{DE} \equiv p_\mathrm{DE}
/ \rho_\mathrm{DE}$, 
might be less than $-1$. In particular, combined data from
WMAP+eCMB+BAO+$H_0$+SNe lead to
\be
\label{p0}
w_\mathrm{DE0}=- 1.17^{+0.13}_{-0.12}\, 
\ee
for a non-constant $w_\mathrm{DE}$
($w_\mathrm{DE}=w_\mathrm{DE0}+w_a(1-a)$
\citep{Chevallier:2000qy,Linder:2002et}), in a flat universe, at $68\%$
confidence level. Note that the Seven-year WAMP+BAO+SNe results had
correspondingly given 
$w_\mathrm{DE0}=-0.93^{+0.12}_{-0.12}$ \citep{Komatsu:2010fb}.

Similarly, for a constant  $w_\mathrm{DE}$ in a flat universe, the
Nine-Year WMAP+eCMB+BAO+$H_0$+SNe data lead to 
\be
\label{p0b}
w_\mathrm{DE}=- 1.084^{+0.063}_{-0.063}\, . 
\ee
This is the improved constraint, following the corresponding ones
of $w_\mathrm{DE}=-0.992^{+0.061}_{-0.062}$
(WMAP+BAO +SNe) of the Five-year WAMP results \citep{Komatsu:2008hk}, and of
$w_\mathrm{DE}=-0.98^{+0.053}_{-0.053}$
(WMAP+BAO+SNe) of the Seven-year WAMP results \citep{Komatsu:2010fb}.

Observing the above sequence of results, we deduce that the increasing
statistics, as well as the increased combinations of data, seem to lead
to a small tendency towards the increasing favoring for the phantom
regime. This can be also observed from the corresponding sequence of
results for a non-flat universe, as well as from different data
combinations.

On the other hand, the standard $\Lambda$CDM model gives, of course,
$w_\mathrm{DE}=-1$, and models of canonical scalar fields lead to
$w_\mathrm{DE}>-1$. If one desires to generate the
$w_\mathrm{DE}<-1$ regime in a scalar field theory in the context of
General Relativity he/she needs a ghost scalar, which leads to several
inconsistencies, especially at the quantum level. However, the above
discussion suggests that we should look at the phantom regime more
thoroughly, since eventually it may be the present state of the universe.
In this letter we summarize briefly the consistent scenarios of
describing the phantom regime, without the need of ghost degrees of
freedom.

\section{Scalar field theory as a model of dark energy}

For completeness, let us very briefly review on a scalar field
$\phi$ with a potential $V(\phi)$ as a scenario of dynamical dark energy
\citep{Ratra:1987rm,Copeland:2006wr}. Such a paradigm results from a
scalar action of the form
\be
\label{p1}
S=\int d^4 x
\sqrt{-g}\left\{\frac{1}{2\kappa^2}R - \frac{1}{2}\partial_\mu
\phi \partial^\mu \phi - V(\phi)\right\}\, ,
\ee
and the dark energy sector is attributed to the scalar field. In
particular, in a spatially flat  Friedmann-Robertson-Walker (FRW) geometry
\be
\label{p2}
ds^2 = - dt^2 + a(t)^2 \sum_{i=1,2,3} \left(dx^i\right)^2\, ,
\ee
the scalar field energy density $\rho_\phi$ and  pressure $p_\phi$ are
respectively given by
\be
\label{p3}
\rho_\phi = \frac{1}{2}{\dot \phi}^2 + V(\phi)\, ,\quad 
p_\phi = \frac{1}{2}{\dot \phi}^2 - V(\phi)\, ,
\ee
and thus the dark energy Equation of State (EoS) parameter $w_\phi$
writes as 
\be
\label{p4}
w_\phi \equiv \frac{p_\phi}{\rho_\phi} 
= \frac{\frac{1}{2}{\dot \phi}^2 - V(\phi)}{\frac{1}{2}{\dot \phi}^2 +
V(\phi)}\, .
\ee
Since usually we assume $V(\phi)\geq 0$, we find $w_\phi > -1$  and
therefore we straightforwardly deduce that the canonical scalar field
(\ref{p1}) cannot describe the phantom 
dark energy where the EoS parameter is less than $-1$. In these lines one
could think of changing by hand the sign of the scalar kinetic term as 
\be
\label{p5}
S=\int d^4 x
\sqrt{-g}\left\{\frac{1}{2\kappa^2}R + \frac{1}{2}\partial_\mu
\varphi \partial^\mu \varphi - \tilde V(\varphi)\right\}\, ,
\ee
which corresponds to the phantom scalar field \citep{Caldwell:1999ew},
since now the EoS parameter $w_\varphi$ is given by
\be
\label{p6}
w_\varphi \equiv \frac{p_\varphi}{\rho_\varphi} 
= \frac{- \frac{1}{2}{\dot \varphi}^2 - \tilde V(\varphi)}
{-\frac{1}{2}{\dot \varphi}^2 + \tilde V(\varphi)}\, , 
\ee
which is less than $-1$. However, we should mention that the negative
kinetic term and the violation of the Null Energy Condition implies that
the energy is unbounded from below at the classical level, while negative
norms
appear at the quantum level\footnote{In order 
to define the ground state, the energy of the quantum system is
indeed bounded from below, but 
there always appear negative norm states, which conflicts the requirement
of   unitarity.
} \citep{Cai:2009zp}. 
The negative norm states generate negative probabilities which conflict
with the usual interpretation of quantum field theory (for example in
\citep{Cline:2003gs} the authors reveal the causality and stability problems
and the possible spontaneous breakdown of the vacuum into phantoms and
conventional particles, arising from the energy
negativity). In Appendix A we discuss the difference with the
ghost appearance in quantum field theory.

From the above it becomes clear that the consistent generation of the
$w_\mathrm{DE}<-1$ regime must arise from scenarios that go beyond the
General-Relativity-based scalar field theory.

\section{Brans-Dicke like model}

In this section we briefly show how the phantom regime may arise in a
Brans-Dicke-type scenario \citep{Elizalde:2004mq}, without the appearance of
any ghost degree of freedom. In the following, for convenience, we define
the effective (total) EoS parameter $w_\mathrm{eff}$ as
\be
\label{p02}
w_\mathrm{eff}\equiv - 1 - \frac{2\dot H}{3H^2} \, ,
\ee
which proves very useful in scenarios where the separation of the total
energy density and pressure into matter and modified gravitational
contributions is difficult
($w_\mathrm{eff}=\Omega_\mathrm{DE}w_\mathrm{DE}$ for a universe with dust
matter).

The action of the Brans-Dicke type model reads \citep{Elizalde:2004mq}:
\be
\label{p7} 
S = \frac{1}{2\kappa^2}\int d^4 x
\sqrt{-g}\,\e^{\alpha\phi}\left\{R - \frac{\gamma}{2} \partial_\mu \phi
\partial^\mu \phi 
- V(\phi)\right\} \, ,
\ee
with $\gamma$ the usual Brans-Dicke parameter and $\alpha$ a constant. 
Therefore, the action in the Einstein frame is given by the scale
transformation 
\be
\label{p8}
g_{\mu\nu}=\e^{- \alpha\phi}g_{E\,\mu\nu}\, ,
\ee
and in the following the subscript $E$ denotes the quantities in the
Einstein frame. Transforming  the action (\ref{p7}) through  (\ref{p8}) we
result to
\begin{eqnarray}
\label{p9} 
&&S = \frac{1}{2\kappa^2}\int d^dx \sqrt{-g_E}\left\{R_E - \left(
\frac{3\alpha^2}{2} 
+ \frac{\gamma}{2}\right)g_E^{\mu\nu} \partial_\mu \phi \partial_\nu \phi 
\right.\nonumber\\
&&\left.\ \ \ \ \ \ \ \ \ \ \ \ \ \ \ \ \ \ \ \ \ \ \ \ \ \ \ \ \ \ \ \ \ \
\ \, - \e^{- \frac{2\alpha}{d-2}\phi}V(\phi)\right\} \, .
\end{eqnarray}
Thus, even if $\gamma$ is negative, in the case where $\frac{3\alpha^2}{2}
+ \frac{\gamma}{2}>0$ the effective kinetic energy of $\phi$ becomes
positive, similarly to the usual scalar field, and therefore the ghost does
not appear. 

Let us provide a specific example, choosing without loss of generality
the potential
\be
\label{p10}
V(\phi) = V_0 \e^{\frac{\phi}{\phi_0}}\, ,
\ee
with constants $V_0$ and $\phi_0$. In this case we find the  
solution \citep{Elizalde:2004mq}
\be
\label{p11}
\phi = - 2\phi_0 \ln\left(\frac{t}{t_0}\right)\, ,\quad 
H = - \frac{\left( \alpha^2 + \gamma \right) \phi_0 + 2 \alpha}
{\alpha^2 \phi_0 \left( \alpha \phi_0 - 1 \right)t}\, ,
\ee
where $t_0$ is an integration constant. 
Then the effective EoS parameter $w_\mathrm{eff}$ from (\ref{p02}) writes
as
\be
\label{p12}
w_\mathrm{eff}= - 1 - \frac{2 \alpha^2 \phi_0 \left( \alpha \phi_0 - 1
\right)}
{3 \left[\left( \alpha^2 + \gamma \right) \phi_0 + 2 \alpha\right]} \, .
\ee
Thus, $w_\mathrm{eff}$ can indeed lie in the
phantom regime if we choose, for example, $\phi_0>0$, $\alpha \phi_0 > 1$
and
$\gamma > 0$\footnote{
In the case of the phantom realization the present universe corresponds to
negative cosmological time $t$.}. We mention here that the above phantom
realization is even more strong than what needed, since not only the dark
energy sector is phantom-like ($w_\mathrm{DE}<-1$), but the total one 
$w_\mathrm{eff}$ lies in the phantom regime too. 

In summary, the  Brans-Dicke type model at hand can generate a phantom
universe, without the presence of a ghost degree of freedom. We stress
that this behavior is not spoiled at the perturbation level, that is the
scenario is free of perturbative instabilities  \citep{Elizalde:2004mq}.

\section{Gauss-Bonnet gravity with a non-minimal scalar field}

Another scenario in the context of modified gravity that may lead to the
realization of the phantom regime without a ghost, is the
scalar-Einstein-Gauss-Bonnet gravity \citep{Nojiri:2005vv,Nojiri:2006je},
which is motivated by string theory. The starting action writes as
\be
\label{p13}
S=\int d^4 x \sqrt{-g}\left\{ \frac{R}{2\kappa^2} - \frac{1}{2}
\partial_\mu \phi \partial^\mu \phi - V(\phi) - \xi(\phi)
\mathcal{G}\right\}\, ,
\ee
where $\mathcal{G}$ is the Gauss-Bonnet combination and $\xi(\phi)$ a
non-minimal coupling function. Variation of the action (\ref{p13}) with
respect to the metric $g_{\mu\nu}$ provides the field equations as follows:
\begin{align}
\label{p14}
0=& \frac{1}{\kappa^2}\left[- R^{\mu\nu} 
+ \frac{1}{2} g^{\mu\nu} R\right]
+ \frac{1}{2}\partial^\mu \phi \partial^\nu \phi - \frac{1}{4}g^{\mu\nu}
\partial_\rho \phi \partial^\rho \phi
\nn
& - \frac{1}{2}g^{\mu\nu}V(\phi) 
+ 2 \left[ \nabla^\mu \nabla^\nu \xi(\phi)\right]R - 2 g^{\mu\nu}
\left[ \nabla^2\xi(\phi)\right]R  \nn
& - 4 \left[
\nabla_\rho \nabla^\mu \xi(\phi)\right]R^{\nu\rho} - 4 \left[
\nabla_\rho \nabla^\nu \xi(\phi)\right]R^{\mu\rho}
\nn
& + 4 \left[ \nabla^2 \xi(\phi) \right]R^{\mu\nu}
+ 4g^{\mu\nu} \left[ \nabla_{\rho} \nabla_\sigma \xi(\phi) \right]
R^{\rho\sigma} \nn
& + 4 \left[\nabla_\rho \nabla_\sigma \xi(\phi) \right] R^{\mu\rho\nu\sigma} 
\, .
\end{align}
Note that in equation (\ref{p14}) the derivatives of curvature, such as
$\nabla R$, do not appear, and therefore derivatives higher than two
do not appear either, which is in contrast with a general $\alpha R^2 +
\beta
R_{\mu\nu}R^{\mu\nu} + \gamma R_{\mu\nu\rho\sigma}R^{\mu\nu\rho\sigma}$
gravity, where fourth derivatives of $g_{\mu\nu}$ appear. Thus,
when we treat the system classically, by specifying the values of
$g_{\mu\nu}$ and $\dot g_{\mu\nu}$ on a space-like hyper-surface 
as initial conditions, the time evolution is uniquely determined.
This situation is similar to the initial conditions in classical
mechanics, 
in which one only needs to specify the values of position and velocity of
the particle. On the other hand, in a general $\alpha R^2 + \beta
R_{\mu\nu}R^{\mu\nu} + \gamma
R_{\mu\nu\rho\sigma}R^{\mu\nu\rho\sigma}$ gravity, 
one needs to specify the values of $\ddot g_{\mu\nu}$ and $\dddot
g_{\mu\nu}$ 
in addition to those of $g_{\mu\nu}$, $\dot g_{\mu\nu}$, in order to obtain
a unique time evolution. 

As a specific example we consider the string-inspired model 
\citep{Nojiri:2005vv}
\be
\label{p15}
V= V_0\e^{-\frac{2\phi}{\phi_0}}\, , \quad \xi(\phi)=\xi_0
\e^{\frac{2\phi}{\phi_0}}\, . 
\ee
Imposing the FRW universe (\ref{p2}) and assuming that the Hubble rate 
is given by $H=h_0/t$, the metric equation  
(\ref{p14}) and the scalar field equation derived from (\ref{p13})
give the following algebraic equations:
\begin{align}
\label{p17}
&V_0 t_1^2 = - \frac{1}{\kappa^2\left(1 + h_0\right)}\left[3h_0^2 \left( 
1 - h_0\right)
+ \frac{\phi_0^2 \kappa^2 \left( 1 - 5 h_0\right)}{2}\right]\\
\label{p18}
&\frac{48 \xi_0 h_0^2}{t_1^2} = - \frac{6}{\kappa^2\left( 1 
+ h_0\right)}\left(h_0 - \frac{\phi_0^2 \kappa^2}{2}\right)\, .
\end{align}
Since $h_0$ is determined at will by suitably choosing $V_0$ and $\xi_0$, 
we can obtain a negative $h_0$, and therefore the effective (total) EoS
parameter in (\ref{p02}) becomes less than $-1$, that is
$w_\mathrm{eff}=-1+2/(3h_0) < -1$,
which corresponds to an effective phantom realization. 
As a numerical example we may choose $h_0=-80/3$, which gives $w= -
1.025$. In this case we find that 
\begin{align}
\label{p19}
V_0t_1^2& = \frac{1}{\kappa^2}\left( \frac{531200}{231}
+ \frac{403}{154}\gamma \phi_0^2 \kappa^2 \right)>0 \nn
\frac{f_0}{t_1^2}& = -\frac{1}{\kappa^2}\left( \frac{9}{49280}
+ \frac{27}{7884800}\gamma \phi_0^2 \kappa^2 \right)\, .
\end{align}

In summary, the scalar-Einstein-Gauss-Bonnet gravity can realize the
phantom regime without a ghost. Finally, note that this scenario is free
of instabilities at the perturbation level
\citep{Koivisto:2006xf,Koivisto:2006ai}. 

\section{$F(R)$ gravity}

In this section, we consider the phantom realization in the context
of $F(R)$ gravity (see
\citep{Nojiri:2003ft,Nojiri:2006ri,Nojiri:2008nt,Nojiri:2010wj}
and references therein). In such a modified gravitational theory the scalar
curvature $R$ in the Einstein-Hilbert action 
is replaced by an appropriate function $F(R)$:
\be
\label{p21}
S_{F(R)}= \int d^4 x \sqrt{-g} \left[ \frac{F(R)}{2\kappa^2} 
\right]\, .
\ee
Alternatively, one can  formulate $F(R)$ gravity in the scalar-tensor
framework. By introducing the auxiliary field $A$, the action (\ref{p21}) 
is rewritten as
\be
\label{p22}
S=\frac{1}{2\kappa^2}\int d^4 x \sqrt{-g} \left\{F'(A)\left(R-A\right) 
+ F(A)\right\}\, .
\ee
Since variation with respect to $A$ gives  $A=R$, substituting
$A=R$ into
the action (\ref{p22})  reproduces the action  (\ref{p21}).
On the other hand, rescaling the metric as 
$g_{\mu\nu}\to \e^\sigma g_{\mu\nu}$ with $\sigma = -\ln F'(A)$, 
the action in the Einstein frame is obtained as follows:
\begin{equation}
\label{p23}
S_E = \frac{1}{2\kappa^2}\int d^4 x \sqrt{-g} \left[ R -
\frac{3}{2}g^{\rho\sigma}
\partial_\rho \sigma \partial_\sigma \sigma - V(\sigma)\right],
\end{equation}
where
\begin{align}
V(\sigma) = &\e^\sigma g\left(\e^{-\sigma}\right)
- \e^{2\sigma} F\left(g\left(\e^{-\sigma}\right)\right) \nn
=& \frac{A}{F'(A)} - \frac{F(A)}{F'(A)^2}\, .
\end{align}
Here the function $g\left(\e^{-\sigma}\right)$ is obtained by solving the
equation
$\sigma =- \ln F'(A)$ in the form of 
$A=g\left(\e^{-\sigma}\right)$.
From expression (\ref{p23}) we deduce that there does not appear ghost
degrees of freedom in $F(R)$ gravity, which is different from the general
higher derivative gravity, with the exception of the
scalar-Einstein-Gauss-Bonnet gravity
of the previous section. 

As a specific example we consider the ansatz $F(R) \propto f_0 R^m$. In
this case the scenario at hand accepts the solution
\be
\label{p24}
H \sim \frac{-\frac{(m-1)(2m-1)}{m-2}}{t}\, ,
\ee
which inserted into (\ref{p02}) leads to effective EoS parameter
\be
\label{JGRG18}
w_\mathrm{eff}=-\frac{6m^2 - 7m - 1}{3(m-1)(2m -1)}\, .
\ee
Thus, when $m>2$ or $1>m>1/2$, we obtain $w_\mathrm{eff}<-1$. Compared with
the Einstein-Hilbert term, the $R^m$ term dominates if $m>2$ when the
curvature is large and if $1>m>1/2$ when the curvature is small. Then the
case $m>2$ might describe the inflation in the early universe and the case
$1>m>1/2$ might correspond to the accelerated expansion of the present
universe. 

In summary, the phantom regime can be realized in $F(R)$ gravity without
ghost degrees of freedom, and the scenario is free of perturbative
instabilities  \citep{Nojiri:2010wj}.

\section{Conclusions}

The Nine-Year WMAP results combined with other cosmological data
\citep{Hinshaw:2012fq} seem to indicate an enhanced favor for the
phantom regime, comparing to previous analyses. This exotic phase
cannot be obtained in the Standard Model of the Universe ($\Lambda$CDM),
or in a General-Relativity-based scalar field theory, and therefore the
above observational results suggest to consider the phantom regime 
more thoroughly. Clearly, easy descriptions such as the consideration of a
by-hand negative kinetic energy of the scalar field cannot lead to
consistent solutions, since the corresponding scenarios would break down at
the quantum level.

Therefore, it seems reasonable that the realization of the phantom regime
without the appearance of ghost degrees of freedom would need a form of
gravitational modification. In this letter we provided three different
scenarios in which this is in principle possible, namely the Brans-Dicke
type gravity, the scalar-Einstein-Gauss-Bonnet gravity, and the $F(R)$
gravity. Furthermore, these scenarios are free of instabilities at the
perturbation level, which is a necessary condition for their validity and
serious consideration (see Appendix B  for some comments on this).
Obviously, one should proceed further, investigating
the corrections to the Newton law, performing the PPN analysis
\citep{Will:2005va} etc, in order to ensure that these scenarios are
consistent with the more accurate solar-system and experimental data. Such
an analysis could provide additional conditions, in order for the above
models to be more realistic (see \citep{Nojiri:2010wj} for general
properties of the above constructions). 

Before closing this work let us make two final comments. Firstly, as it is
well known one can perform conformal transformations from the Jordan to the
Einstein frame, that is from a ``modified gravity'' action to a 
canonical 
``scalar field'' one \citep{Capozziello:2011et}. 
In general a phantom universe may result to a
finite-time future singularity called ``Big Rip'' singularity 
but it is well-known that this kind of singularity does not occur in the 
model of canonical scalar field. The finite time where the singularity occurs in the 
modified gravity is transformed into the infinite future in the canonical scalar field 
model by the conformal transformation.

In conclusion, the increasing favoring of the phantom regime, as long as it
will not
be reversed in the future combined observational dark-energy constraints,
may serve as a strong indication towards gravity modification, in a
similar way that the (non-phantom) universe acceleration established the
cosmological constant in standard cosmology.

\acknowledgments
This work started at ``3rd International Workshop on
Dark Matter, Dark Energy and Matter-Antimatter Asymmetry'' 
at NTHU (Hsinchu) and NTU (Taipei) in Taiwan. 
We are indebted to all the organizers, especially Prof. C.-Q. Geng, 
for the hospitalities and giving a chance for this collaboration. 
The work by SN is supported in part by Global COE Program of Nagoya
University 
(G07) provided by the Ministry of Education, Culture, Sports, Science \&
Technology and by the JSPS Grant-in-Aid for Scientific Research (S) \#
22224003
and (C) \# 23540296.
ENS thanks the
National Center for Theoretical Sciences, Hsinchu, Taiwan for
warm hospitality during the preparation of this work. His research is
implemented within the framework of the Operational Program ``Education and
Lifelong Learning'' (Action's Beneficiary: General Secretariat for Research
and Technology), and is co-financed by the European Social Fund (ESF) and
the Greek State.

\section*{Appendix A: Ghost and negative norm state in quantum field theory}
\label{AppA}

In   classical field theory  the energy density of the ghost field is
unbounded from below. Then  one could say that  if any ghost field exists
then the vacuum will decay to a  pair-creation of the ghost particles. As
long as we know, of course, no ghost field has been discovered in nature. 
The ghost fields, however, appear in unphysical sectors in gauge theory,
string theory, etc, when we quantize the system Lorentz-covariantly. In
these field theories the energy is surely bounded from below and therefore
the vacuum {\it never} decays. Additionally, apart from of the unbounded
energy, the ghost field generates negative norm states. 
These negative states are combined with the states generated 
by the unphysical modes, like the longitudinal mode, and eventually only
positive and zero norm states appear in the physical space defined by the
BRS charge \citep{Kugo:1977yx}. 
If the negative norm states could appear in the physical space then
negative probability would be generated, 
which conflicts with the Copenhagen interpretation of the wave functions
$\psi$ 
corresponding to quantum states, where the norm $\psi^\dagger \psi$  of the
wave function 
can be regarded as a probability. 

As a simple example, we consider the system composed by the oscillating
modes 
$\left\{ \alpha^\mu_n \right\}$, $\left\{ c_n \right\}$, and $\left\{ b_n
\right\}$ 
of the string coordinates in $d$ dimensions, ghost, and anti-ghost
respectively. 
Here $\mu=0,1,2,\cdots,d-1$ and $n$ is an integer, but we now omit the zero
modes corresponding to $n=0$ 
since these modes are irrelevant in the arguments here, although they have
rich structures 
in string theories. 
The hermiticities of these oscillating modes are assigned as follows:
\be
\label{A0}
{\alpha^\mu_n}^\dagger = \alpha^\mu_{-n}\, ,\quad c_n^\dagger 
= c_{-n} \, ,\quad b_n^\dagger = b_{-n} \, .
\ee
These oscillating modes satisfy the following (anti-) commutation relations:
\be
\label{A1}
\left[ \alpha^\mu_n,\, \alpha^\nu_m \right] = n \eta^{\mu\nu}\delta_{0,n+m}\,
,\quad 
\left\{ b_n, c_n \right\} = n\delta_{0,n+m}\, ,
\ee
with $\left[ \ , \ \right]$ and $\left\{\ ,\ \right\}$ denoting commutator
and 
anti-commutator, respectively. 

The Hamiltonian $H$ is given by 
\be
\label{A2}
H = \sum_{n=1}^\infty \left\{ \sum_{\mu=0}^{d-1}\alpha^\mu_{-n} \alpha^\mu
_n 
+ b_{-n} c_n + c_{-n} b_n \right\}\, ,
\ee
and the commutation relations between the Hamiltonian $H$ and the
oscillating modes are 
given by
\be
\label{A3}
\left[ H, \alpha^\mu_n \right] = - n \alpha^\mu_n\, ,\quad 
\left[ H, c_n \right] = - n c_n\, ,\quad 
\left[ H, b_n \right] = - n b_n\, .
\ee
Therefore if we define the vacuum state $\left| 0 \right>$ by 
\be
\label{A4}
\alpha_n \left| 0 \right> = c_n \left| 0 \right> = b_n \left| 0 \right> =
0\, 
\ee
for positive $n$, the energy in the Fock space given by multiplying the
vacuum state 
$\left| 0 \right>$ with $\left\{ \alpha^\mu_n \right\}$, $\left\{ c_n
\right\}$ and $\left\{ b_n \right\}$, 
with negative $n$ is positive semi-definite. 

At this point we should mention that there appear negative norm states. 
Firstly we may assume that the vacuum state has positive norm and that it
can be normalized to be unity 
$\left< 0 | 0 \right> = 1$. 
If we consider the following states, for example, 
\begin{align}
\label{A5}
\left|n, \alpha^0\right> \equiv& \frac{1}{\sqrt{n}} \alpha^0_{-n} \left| 0
\right>\, \nn
\left|n, b-c\right> \equiv& \frac{1}{\sqrt{2n}} \left(b_{-n} - c_{-n} \right)
\left| 0 \right>\, ,
\end{align}
then
these states have negative norms:
\be
\label{A6}
\left<n, \alpha^0 | n, \alpha^0 \right> = \left<n, b-c|n, b-c\right>=-1\, .
\ee
However, note that these negative norm states only appear as a combination
of the zero 
norm states in the physical space defined by the BRS charge
\citep{Kato:1982im} (see 
\citep{Ito:1985qa} for the case of superstrings based on
Neveu-Schwarz-Ramond Model). 

In summary, in the known framework of quantum field theories there can
appear negative norm states, but the energy of the system, including
ghosts, is positive semi-definite. Thus, the vacuum {\it never} decays in 
quantum field theory. 

\section*{Appendix B: Ghost in higher derivative models}
\label{AppB}

We now briefly show that   higher derivative constructions
 contain ghost in
general. 
As an example we may consider the following model:
\be
\label{A7}
S = \int d^4 x \sqrt{-g } \left( \Box \phi \right)^2\, ,
\ee
where $\phi$ is a scalar field and $\Box$ is the d'Alembertian. 
By introducing an auxiliary field $\zeta$, the action (\ref{A7}) can be
rewritten in the 
following form:
\begin{align}
\label{A8}
S =& \int d^4 x \sqrt{-g} \left( 2 \zeta^2 - \zeta \Box \phi \right) \nn
=& \int d^4 x \sqrt{-g} \left( 2 \zeta^2 + \partial_\mu \zeta \partial^\mu
\phi \right) \, .
\end{align}
By defining new fields $\varphi_\pm$ by
\be
\label{A9}
\zeta = \frac{\varphi_+ + \varphi_-}{\sqrt{2}}\, ,\quad 
\phi  = \frac{\varphi_+ - \varphi_-}{\sqrt{2}}\, ,
\ee
the action is further rewritten as
\begin{align}
\label{align}
S =& \int d^4 x \sqrt{-g} \left\{ \left( \varphi_+ + \varphi_-\right)^2 
+ \frac{1}{2}\partial_\mu \varphi_+ \partial^\mu \varphi_+ \right. \nn
& \left.  - \frac{1}{2}\partial_\mu \varphi_- \partial^\mu \varphi_-
\right\} \, ,
\end{align}
which implies that $\varphi_+$ is a ghost. 

However, in suitably constructed higher derivative scenarios, such is the
Galileon one \citep{Nicolis:2008in}, the background equations of motion do
not contain  higher than second derivatives, and thus ghost do not appear
at this level. But this is not a proof
that ghosts cannot appear  by the canonical formalism (it is a necessary
but not sufficient condition), since   they can appear at the
perturbation level, directly or indirectly (as superluminal propagation).
This proves to be the case in  Ho\v{r}ava-Lifshitz gravity
\citep{Bogdanos:2009uj} as well as in non-linear massive gravity
\citep{Deser:2012qx}.

 Therefore, we deduce that the complete investigation
of the ghost subject is a crucial requirement for the acceptance of a
theory, especially if it allows for the phantom realization. We thus
stress that not all higher derivative theories  lead to the
appearance of ghosts, and as we have shown $F(R)$ or Gauss-Bonnet gravity
do not contain ghosts at the background or perturbation levels.

\end{document}